\documentstyle[12pt,epsfig]{article}
\begin{document}

\begin{minipage}{12cm}
\hspace*{8.7cm} {\bf Preprint JINR}\\
\hspace*{9.3cm}{\bf E2-2005-4}\\
\hspace*{9.0cm}{\bf Dubna, 2005}
\end{minipage}
\vskip 3cm

\begin{center}
{\bf
Z-SCALING AND  INCLUSIVE PARTICLE PRODUCTION\\[0.1cm]
 IN  $pp$ \  COLLISIONS  AT  HIGH-$p_T$
}

\vskip 5mm
M.V. Tokarev $^{a,}$\footnote{E-mail: tokarev@sunhe.jinr.ru},
Yu.A. Panebratsev $^{a,}$\footnote{E-mail: panebrat@sunhe.jinr.ru},
G.P. \v {S}koro $^{b,}$\footnote{E-mail: goran@ff.bg.ac.yu},
and  I. Zborovsk\'{y} $^{c,}$\footnote{E-mail: zborovsk@ujf.cas.cz}

\vskip 5mm

{\small

(a) {\it Veksler and Baldin Laboratory of High Energies,\\
Joint Institute for Nuclear Research,\\
141980, Dubna, Moscow region, Russia}

\vskip 0.2cm

(b) { \it Institute of Nuclear Sciences "Vin\v {c}a",\\
Faculty of Physics, University of Belgrade,\\
Belgrade, Serbia and Montenegro}

\vskip 0.2cm

(c)  {\it Nuclear Physics Institute,\\
Academy of Sciences of the Czech Republic, \\
\v {R}e\v {z}, Czech Republic}
}

\end{center}

\vskip 10mm

\begin{center}
\begin{minipage}{150mm}
\centerline{\bf Abstract}
The properties of particle production in $pp$ collisions are basis
for analysis of $pA$ and $AA$  interactions and verification of theory.
We analyzed  inclusive particle production using $z$-scaling established
in proton-proton and proton-antiproton collisions at the U70, ISR, Tevatron and RHIC.
This scaling reflects symmetry of hadron structure,
interaction and particle formation.
It is shown that new data on charged hadron and $\pi^0$-meson
spectra obtained in  $pp$ collisions at the Relativistic Heavy Ion Collider
 confirm the $z$-scaling.
\end{minipage}
\end{center}

\vskip 30mm

\begin{center}
  Presented by  I.Zborovsk\'{y} at the XVII International Conference on Ultra
Relativistic Nucleus-Nucleus Collisions "Quark Matter'2004"
January 11-17, 2004, Oakland, USA.
\end{center}

\newpage
{\section{Introduction}}

With advent of Relativistic Heavy Ion Collider at BNL (USA) new era of
tremendous accumulation of data in high energy domain  began.
Excellent  possibilities to search for new physics phenomena and verification
of theory are opened.

Particle production with high transverse momenta is of specific interest.
 It is traditionally  connected with local character of hadron interactions
which are  believed  to have adequate  description in  the theory of QCD.
Locality of the interaction is expressed in terms of the hadron
  constituents.
Scaling features of high-$p_T$ hadron spectra reflects
self-similarity of the constituent interactions.
Therefore, search and verification of
scaling regularities  at RHIC energy domain are important.
This would  stimulate more profound investigations  of particle
production at RHIC and later at LHC.

Up to date, the investigation of hadron properties in the
high energy collisions has revealed widely known scaling laws.
From the most popular and famous, let us mention the Feynman
scaling \cite{Feynman} observed in inclusive particle
production, the Bjorken scaling \cite{Bjorken}
in deep inelastic scattering (DIS), $y$-scaling \cite{Bosted}
valid in DIS on
nuclei, limiting fragmentation established for nuclei
\cite{Benecke}, scaling behaviour of the cumulative particle
production \cite{Baldin,Stavinsky,Leksin},
the Koba-Nielsen-Olesen (KNO) scaling
\cite{KNO}, and others.
Another expression of self-similarity in hadronic interactions
is described by quark counting rules \cite{Brodsky}.
All these scalings have restricted range of validity.
The established  violations indicate
manifestation of dynamical mechanisms of constituent interactions
in the region beyond  their applicability.
In particular, this led to discovery of QCD evolution in DIS.

New method to study the properties of particle structure,
constituent interactions and particle formation is $z$-scaling
\cite{Zscal}.
The scaling was observed in proton-(anti)proton
high energy collisions at
U70, ISR, S$\bar{\rm p}$pS and Tevatron.
In this paper we verify  $z$-scaling using new data on
 charged hadron and $\pi^0$-meson
spectra obtained in  $pp$ collisions at the Relativistic Heavy Ion Collider.

\vskip 0.5cm
  {\section{Basic principles of $z$-scaling}}

 The idea of $z$-scaling is based on the principles of locality,
self-similarity and fractality of hadronic interactions.
The principles reflect structure of colliding objects, features of
the underlying constituent interactions and mechanism of
particle production.
Their most pronounced manifestation is
expected in the inclusive production of particles with large
transverse momenta.

{\subsection{Locality principle}}

Locality of hadron interactions is considered at constituent level.
We  assume that gross features of the inclusive particle
  distributions for the reaction
  \begin{equation}
  M_{1}+M_{2} \rightarrow m_1 + X
  \label{eq:r1}
  \end{equation}
can be described  in  terms of the constituent
sub-process
\begin{equation}
(x_{1}M_{1}) + (x_{2}M_{2}) \rightarrow m_{1} +
(x_{1}M_{1}+x_{2}M_{2} + m_{2})
\label{eq:r2}
\end{equation}
satisfying to the condition
\begin{equation}
(x_{1}P_{1} + x_{2}P_{2} - p)^{2} = (x_{1}M_{1} + x_{2}M_{2} +
m_{2})^{2}.
\label{eq:r3}
\end{equation}
The $x_{1}$ and $x_{2}$ are fractions
 of the incoming four momenta $P_1$ and $P_2$ of the colliding
objects with the masses $M_1$ and $M_2$. The $p$ is four-momentum of
the inclusive particle with the mass $m_1$.
The parameter $m_{2}$ is introduced to satisfy the internal
conservation laws (for isospin, baryon number, strangeness, and
so on).
The equation (\ref{eq:r3}) describes the energy-momentum conservation
for the elementary constituent sub-process.

  \vskip 0.5cm
  {\subsection{Self-similarity principle}}

Self-similarity is  scale-invariant property connected with
dropping of certain dimensional quantities out of physical
picture of the interaction.
Self-similarity parameters are constructed as specific combinations of
these quantities. The scaling function $\psi(z)$ depends in a self-similar manner
on a single variable $z$.
The  variable $z$  is dimensionless combination of quantities
which characterize particle production in high energy inclusive reactions.
It depends on momenta and masses of colliding and inclusive particles,
multiplicity density and fractal dimensions of the incident objects.
The scaling function is expressed via the invariant differential
cross section $Ed^3\sigma/dp^3$ as follows
\begin{equation}
 \psi(z) = - \frac{\pi s}{(dN/d\eta) \sigma_{in}}J^{-1}
 E\frac{d^3\sigma}{dp^{3}}.
 \label{eq:r4}
 \end{equation}
Here $s$ is the collision center-of-mass energy squared,
$\sigma_{in}$ is the inelastic cross section, $J$ is the
corresponding Jacobian, and $dN/d\eta$ is the particle
multiplicity density.
The function $\psi(z)$ is normalized as
\begin{equation}
\int_{0}^{\infty} \psi(z) dz = 1.
\label{eq:b6}
\end{equation}
The relation allows us to give the physical meaning
of the scaling function $\psi(z)$ as  probability density to form
a particle  with the corresponding value of the variable $z$.

  \vskip 0.5cm
  {\subsection{Fractality principle}}

Principle of fractality states that variables used in the
description of the process diverge in terms of the resolution
\cite{Mandelbrot,Nottale}.
This property is characteristic for the scaling variable
\begin{equation}
z = z_0 \Omega^{-1},
\label{eq:r5}
\end{equation}
where
\begin{equation}
\Omega(x_1,x_2)=(1-x_1)^{\delta_1}(1-x_2)^{\delta_2} .
\label{eq:r6}
\end{equation}
The variable $z$ has character of a fractal measure.
For a given production process (\ref{eq:r1}),
its finite part $z_0$ is the ratio
of the transverse energy released in the
binary collision of constituents (\ref{eq:r2})
and the average multiplicity density $dN/d\eta|_{\eta=0}$.
The divergent part
$\Omega^{-1}$ describes the resolution at which the collision of
the constituents can be singled out of this process.
The $\Omega(x_1,x_2)$ represents relative number of all initial
configurations containing the constituents which carry fractions
$x_1$ and $x_2$ of the incoming momenta.
The $\delta_1$ and $\delta_2$ are the anomalous fractal
dimensions of the colliding objects (hadrons or nuclei).
The momentum fractions $x_1$ and $x_2$ are determined in a way to
minimize the resolution $\Omega^{-1}(x_1,x_2)$ of the fractal
measure $z$ with respect to all possible sub-processes
(\ref{eq:r2}) subjected to the condition (\ref{eq:r3}).
The variable $z$ was interpreted \cite{Zscal} as a particle
formation length.

 \vskip 0.5cm
{\section{Ingredients and properties of $z$-scaling}}

One of the ingredients in the definition of the variable $z$
is the particle multiplicity density $ dN/d\eta|_{\eta=0}$ in central
region of collision.
We have used the experimental data for this quantity in our analysis.
Another  ingredients  are the anomalous
fractal dimensions $\delta$ of the colliding objects.
It was found that,
 for the  nucleon-nucleon collisions, $\delta$ does not depend on the
 colliding energy $\sqrt s $ and other kinematical variables ($p_T, \theta$).

Using experimental data on high-$p_T$ particle production in $pp$ and $\bar pp$
collisions at U70, ISR, S$\bar{\rm p}$pS and Tevatron, we have established
energy and angular independence of the scaling function $\psi(z)$ for the same value of
$\delta$. In the region
of large $z$, the scaling function reveals power behavior, $\psi(z) \sim z^{-\beta}$.
The asymptotic behavior of $\psi(z)$  reflects self-similarity and fractality
of hadron interactions at small scales.

\vskip 0.5cm
{\section{First verification of $z$-scaling  at RHIC}}

In present analysis of new  RHIC data
we verify the  energy independence of the scaling function $\psi(z)$
in $pp$ collisions  at $\sqrt s =200$~GeV.
We have used the data on inclusive cross sections of charged
hadron \cite{Adams} and $\pi^0$-meson \cite{Phenix} production
obtained by the STAR and PHENIX collaborations, respectively.
Inclusive spectra were measured in the central region of collision
and cover the transverse momentum range up to $p_T=13$~GeV/c.

 The invariant differential cross section for charged hadrons produced in $pp$
 collisions  as a function of the transverse
 momentum  is shown in Figure 1(a).
 Different symbols correspond to data at different colliding energies
 $\sqrt s = 11.5-200$~GeV.
    The experimental data
    demonstrate  strong energy dependence of
the cross sections which increases with $p_T$.
The energy dependence of the cross sections is contrasted with
the energy independence of $z$-presentation of the data shown in Figure 1(b).
Data measured by the STAR collaboration cover the range of $z$ from 0.3 up to 10.
This includes asymptotic power regime which starts approximately  at $z>4$.
The result shows that the STAR data  confirm  the energy  independence of the $z$-scaling
for charged particles produced in $pp$ collisions.

The invariant differential cross section for neutral pions  produced in $pp$
 collisions  as a function of the transverse
 momentum   is shown in Figure 2(a).
The plotted data from ISR were obtained at the energies  $\sqrt s =23-62$~GeV.
The PHENIX data from RHIC were measured at $\sqrt s = 200$~GeV.
Both data sets correspond to the central rapidity region.
The differential cross sections strongly depend on the collision energy
in the similar way as for the charged particles.
The corresponding $z$-presentation of data is shown in Figure 2(b).
The data measured by the PHENIX collaboration cover the range of $z$ from 1 up to 13.
One can see that these data are in  good agreement with $z$-scaling  for
$\pi^0$-meson production obtained at ISR energies. The power asymptotic
regime is clear observed in this case as well.
The result shows that the  PHENIX data confirm  the energy  independence of the $z$-scaling
for neutral mesons  produced in $pp$ collisions.

\vskip  0.5cm
{\section{Conclusions}}

$Z$-scaling  is  specific feature of inclusive high-$p_T$ particle production
   established in  proton-proton and proton-antiproton collisions.
The scaling  was observed in numerous  data obtained
at U70, ISR, S$\bar{\rm p}$pS and Tevatron.

In this paper we exploit new
data on inclusive production of charged hadrons and $\pi^0$-mesons measured
 by the STAR and PHENIX collaborations to verify the $z$-scaling.
We have shown that these data confirm the energy independence
 of the scaling function $\psi(z)$.
The STAR and PHENIX data allow to test asymptotic regime of the scaling function
for $z>4$.
They confirm  that asymptotic regime of $\psi(z)$ at high-$z$
is governed by a power law, $\psi(z)\sim z^{-\beta}$.
The slope parameter
$\beta$ is  independent of  center-of-mass energy $\sqrt s$ and transverse
momentum  $p_T$ over a wide kinematical range.
This is consequence of constant value of $\delta$
characterizing nucleon fractal structure at different scales.
In conclusion, the RHIC data confirm $z$-scaling in the range  of
$\sqrt s = 11.5-200$~GeV and $p_T = 0.5-13$~GeV/c.
We consider that  established properties of data $z$-presentation
could give additional constraints
on phenomenological quantities (parton distribution and fragmentation functions etc.)
in calculations of hard processes using QCD.

The obtained results demonstrate  that
self-similarity and fractality are  features of hadron structure,
interactions and particle formation  which reflect specific symmetries in Nature.
$Z$-scaling  can be used   as a tool to
search for new phenomena using  hard probed such as  high-$p_T$ hadrons,
direct photons, heavy quarkonia  and  jets at the RHIC,  Tevatron and LHC.

\vskip 1cm {\bf Acknowledgments.} This work has been partially
supported by the IRP AVOZ10480505 and by the  Grant Agency of the
Czech Republic under the contract No. 202/04/0793. We are grateful
to M.\v{S}umbera for support of these investigations.


\newpage
\begin{minipage}{4cm}

\end{minipage}

\vskip 4cm
\begin{center}
\hspace*{-2.5cm}
\parbox{5cm}{\epsfxsize=5.cm\epsfysize=5.cm\epsfbox[95 95 400 400]
{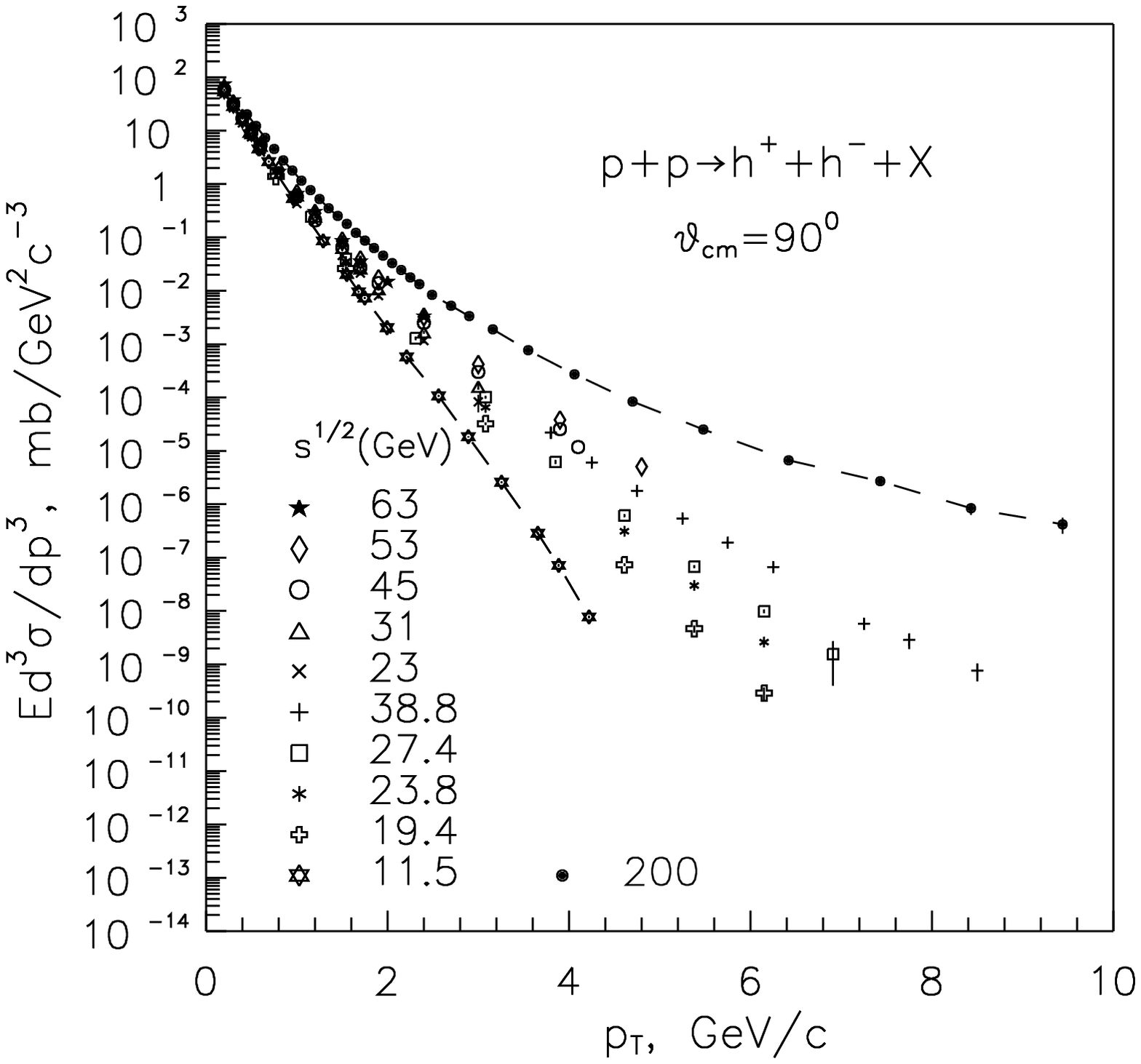}{}} \hspace*{3cm}
\parbox{5cm}{\epsfxsize=5.cm\epsfysize=5.cm\epsfbox[95 95 400 400]
{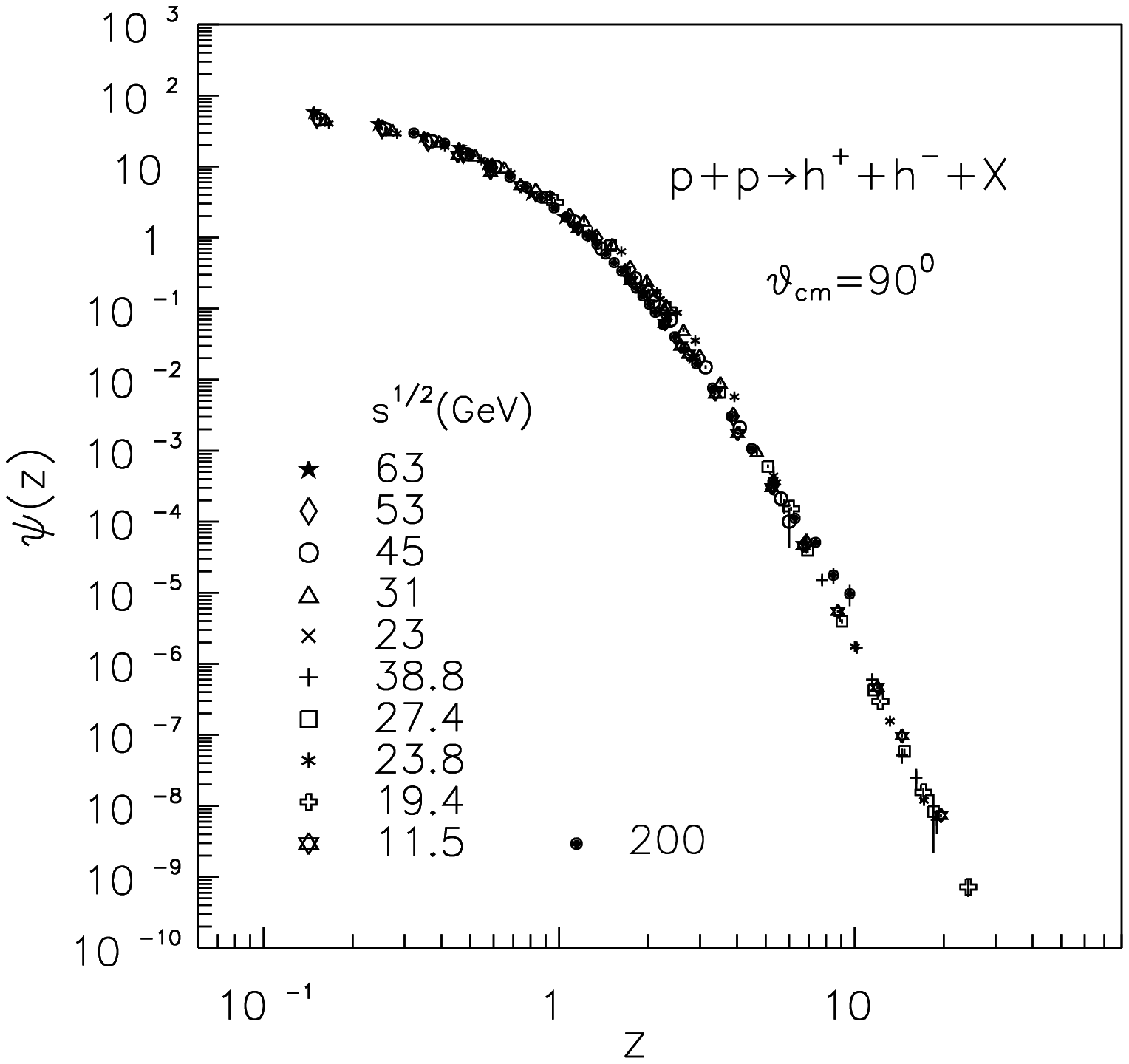}{}} \vskip -0.5cm
\hspace*{0.cm} a) \hspace*{8.cm} b)\\[0.5cm]
\end{center}

{\bf Figure 1.}
(a) Inclusive  cross sections
  of charged hadrons produced in  $pp$ collisions
at $\sqrt s =11.5-63$ and 200~GeV for  $\theta_{cm} \simeq 90^{0}$
as  functions of the transverse momentum.
Experimental  data are taken from
 \cite{Protvino}-\cite{Alper}
  and \cite{Adams}.
(b) $Z$-presentation of the same data.

\vskip 5cm

\begin{center}
\hspace*{-2.5cm}
\parbox{5cm}{\epsfxsize=5.cm\epsfysize=5.cm\epsfbox[95 95 400 400]
{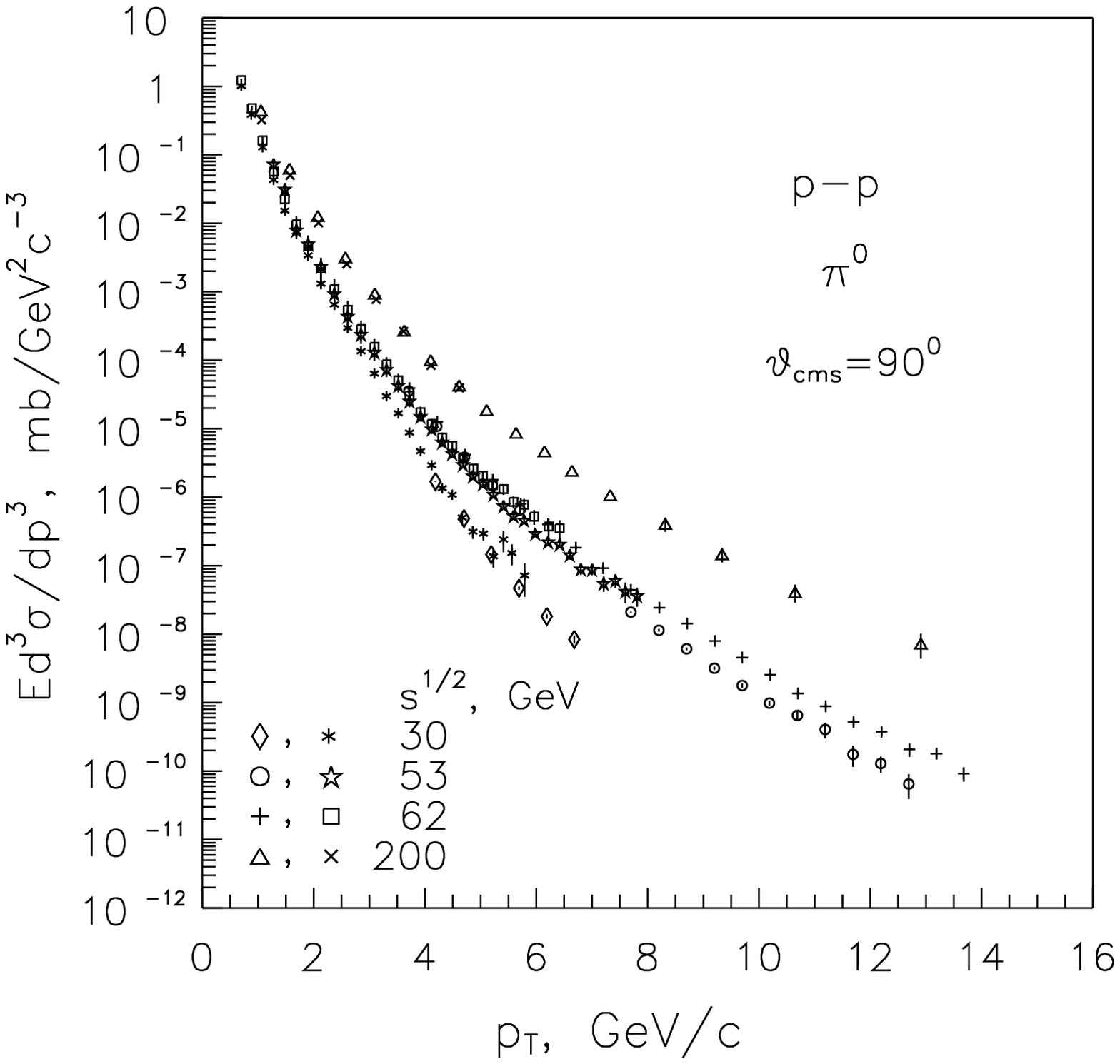}{}} \hspace*{3cm}
\parbox{5cm}{\epsfxsize=5.cm\epsfysize=5.cm\epsfbox[95 95 400 400]
{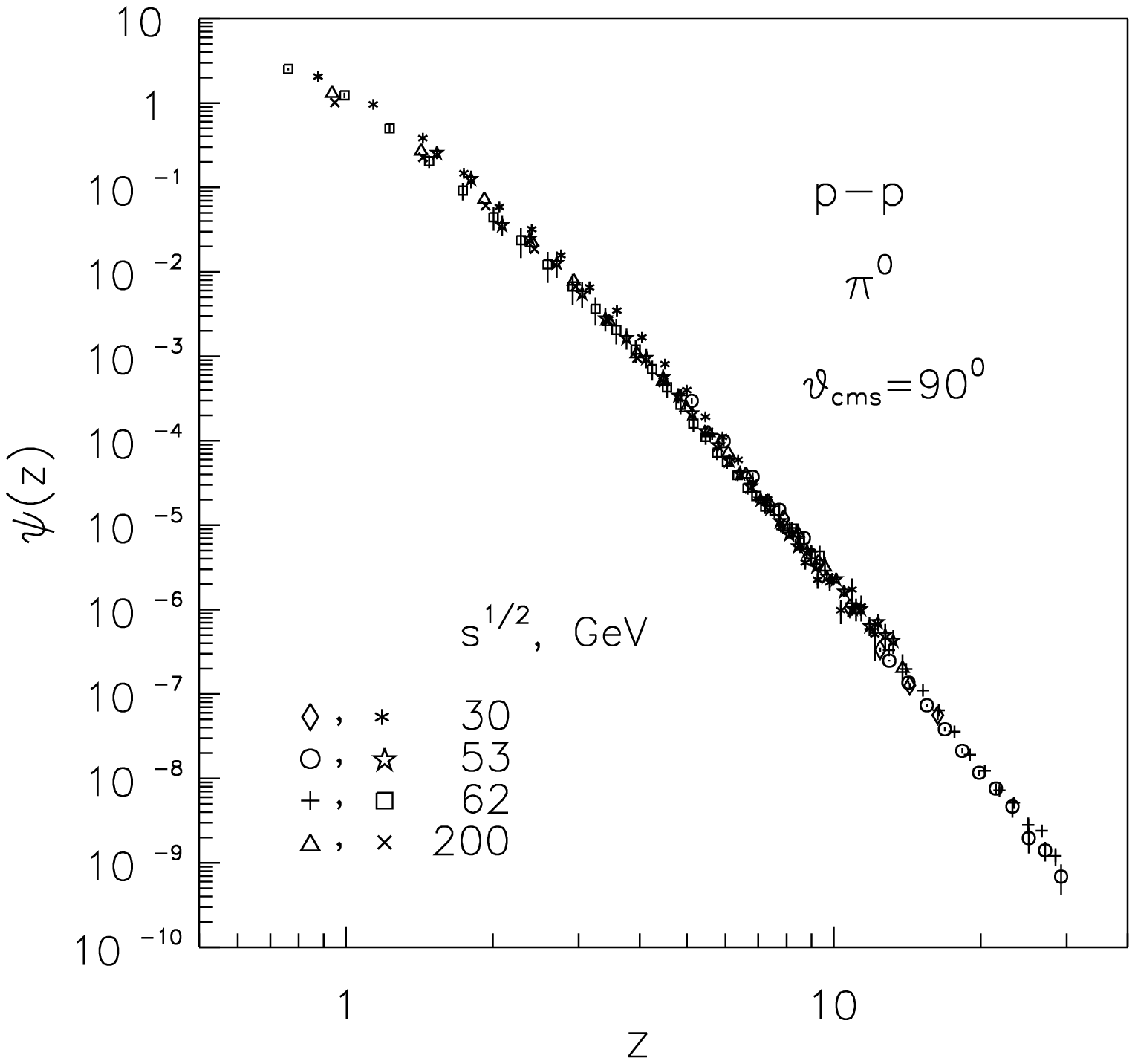}{}} \vskip -1.cm
\hspace*{0.cm} a) \hspace*{8.cm} b)\\[0.5cm]
\end{center}

{\bf Figure 2.}
The dependence of  the inclusive cross section of $\pi^0$-meson production
on the transverse
momentum  in $pp$ collisions at $\sqrt s = 30,53,62$ and 200~GeV
for the angle $\theta_{cm} \simeq 90^0$.
The experimental data  are taken from
\cite{Angel}-\cite{Eggert}
and \cite{Phenix}.
(b) The corresponding scaling function $\psi(z)$.

\vskip 0.5cm

\end{document}